\documentclass[aps,amsmath,showpacs,amsfonts,10pt]{revtex4}
\usepackage{epsfig,graphicx}
\usepackage[english]{babel}
\usepackage{amsfonts}
\usepackage{amsmath}
\usepackage{latexsym}
\usepackage{graphics,bm}
\usepackage{subfigure}
\usepackage{dcolumn}
\usepackage{bm}
\usepackage{rotating}

\begin{document}

\title{Coherence Properties of a Bose-Einstein Condensate in an Optical Superlattice}

\author{Aranya B Bhattacherjee}
\affiliation{Max Planck-Institute f\"ur Physik komplexer Systeme, N\"othnitzer Str.38,01187 Dresden,Germany \\ and \\ Department of Physics, Atma Ram Sanatan Dharama College, University of Delhi (South Campus), Dhaula Kuan, New Delhi-110021, India. }

\begin{abstract}
We study the effect of a one dimensional optical superlattice on the superfluid fraction, number squeezing, dynamic structure factor and the quasi-momentum distribution of the Mott-insulator. We show that due to the secondary lattice,there is a decrease in the superfluid fraction and the number fluctuation. The dynamic structure factor which can be measured by Bragg spectroscopy is also suppressed due to the addition of the secondary lattice. The visibility of the interference pattern (the quasi-momentum distribution)of the Mott-insulator is found to decrease due to the presence of the secondary lattice. Our results have important implications in atom interferometry and quantum computation in optical lattices.
\end{abstract}


\maketitle

\section{Introduction}

When a gas of ultracold atoms is loaded into an optical lattice,its properties are modified strongly\cite{Morsch06}. Ultracold bosons trapped in such periodic potentials have been widely used recently as a model system for the study of some fundamental concepts of quantum physics like Josephson effects\cite{Anderson98}, squeezed states,\cite{Orzel01} landau-Zener tunneling and Bloch oscillations  \cite{Morsch01}and superfluid-Mott insulator transition \cite{Greiner02}.
Using superposition of optical lattices with different periods \cite{Peil03}, it is now possible to generate more sophisticated periodic potentials characterized by a richer spatial modulation, the so-called optical superlattices. An important and exciting application of optical superlattice is quantum computation \cite{Sebby06}.  The physics of one-dimensional optical superlattices has been a subject of recent research, including fractional filling Mott insulator (MI) domains \cite{Bounsante04}, dark \cite{Louis04}and gap  \cite{Louis05} solitons, the Mott-Peierls transition \cite{Dimtrieva68}, non-mean field effects \cite{Rey04}, phase-diagram in two colour superlattices \cite{Roth03}, Bloch-Zener and dipole oscillations \cite{Breid07}, collective oscillations  \cite{Chun05}and Bloch and Bogoluibov spectrum \cite{Bhattacherjee07}.
A key observable in these systems is the interference pattern observed after releasing the gas from the lattice and letting it expand for a certain time of flight. Monitoring the evolution of this interference pattern reveals e g., the superfluid fraction, number squeezed states \cite{Orzel01,Hadzibabic04}, quasi-momentum distribution, observation of collapse and revivals of coherence due to atomic coherence \cite{Mandel02} and superfluid to Mott insulator transition \cite{Greiner02,Zwerger03}. Further coherence properties of Bose-Einstein condensates offer the potential for improved interferometric phase contrast. The MI state plays a central role for various quantum information processing schemes \cite{Rabl03}. Because of the experimental importance of BEC in optical lattices, it is crucial to understand the influence of the secondary lattice which is emerging as a new manipulating tool on the coherence properties of a BEC. In the present paper, we study in what way the superfluid fraction, number fluctuation, the dynamic structure factor and the quasi-momentum distribution (and hence the visibility of the interference pattern) of the MI is influenced by the addition of the secondary lattice.

\section{The Bogoluibov approximation to the Bose-Hubbard Hamiltonian}
The light shifted potential of the superlattice is described as

\begin{equation}
V(z)=V_{1}\cos^{2}\left(\frac{\pi z}{d_{1}} \right)+V_{2}\cos^{2}\left(\frac{\pi z}{d_{2}}+\phi \right)  
\end{equation}

Here $d_{1}$ and $d_{2}$ are respectively, the primary and secondary lattice constants. $V_{1}$ and $V_{2}$ are the respective amplitudes. The secondary lattice acts as a perturbation and hence we will take $V_{2}<<V_{1}$.  $\phi$ is the phase of the secondary lattice. When $\phi=0$, each site of the lattice is perfectly equivalent due to the symmetries of the system so that the population and on site energies are same at each site. An asymmetry is introduced when $\phi\not=0$ and hence the onsite energies are not the same at each site. 
The harmonic trapping potential is given by $V_{ho}(r,z)=\frac{m}{2}\left(\omega^{2}_{r}r^{2}+\omega^{2}_{z}z^{2}\right)$  and the optical superlattice potential is given as $V_{op}=E_{R}\left(s_{1}\cos^{2}(\frac{\pi z}{d})+s_{2}\cos^{2}(\frac{\pi z}{2d}\right)$. In our case we take $d_{2}=2d_{1}=2d$ which gives rise to a periodic double well potential. Also $s_{1}$ and $s_{2}$ are the dimensionless amplitudes of the primary and secondary superlattice potentials with $s_{1}>s_{2}$. $E_{R}=\frac{\hbar^2\pi^2}{2md^2}$ is the recoil energy ($\omega_{R}=\frac{E_{R}}{\hbar}$ is the corresponding recoil frequency) of the primary lattice.$U=\frac{4\pi a\hbar^2}{m}$ is the strength of the two body interaction and $a$ is the two body scattering length.We take $\omega_{r}>\omega_{z}$ so that an elongated cigar shaped BEC is formed. The harmonic oscillator frequency corresponding to small motion about the minima of the optical superlattice is $\omega_{s}\approx \frac{\sqrt{s_{1}}\hbar \pi^2}{md^2}$. The BEC is initially loaded into the primary lattice and the secondary lattice is switched on slowly so that the BEC stays in the vibrational ground state. The frequency of each minima of the primary lattice is not perturbed significantly by the addition of the secondary lattice. $\omega_{s}>>\omega_{z}$ so that the optical lattice dominates the harmonic potential along the $z$-direction and hence the harmonic potential is neglected. Moreover we also take a sufficiently large harmonic confinement in the $xy$ plane which effectively reduces the problem to one-dimension. The strong laser intensity will give rise to an array of several quasi-two dimensional pancake shaped condensates.Because of the quantum tunneling, the overlap between the wavefunctions between two consecutive layers can be sufficient to ensure full coherence. Following our earlier work \cite{Bhattacherjee07} the effective one-dimensional Bose-Hubbard Hamiltonian for $I$ lattice sites and $\phi\not=0$ is written as

\begin{equation}
H=-\sum_{j} J_{j} \left[\hat{a}_{j}^{\dagger}\hat{a}_{j+1}+\hat{a}_{j+1}^{\dagger}\hat{a}_{j}\right]
+\frac{U'_{eff}}{2}\sum_{j}\hat{a}_{j}^{\dagger}\hat{a}_{j}^{\dagger}\hat{a}_{j}\hat{a}_{j}
+\sum_{j}\epsilon_{j}\hat{a}_{j}^{\dagger}\hat{a}_{j}.
\end{equation}

Here $J_{j}$ is the site dependent strength of the Josephson coupling and is different when going from $j-1$ to $j$ and $j$ to $j+1$.The two Josephson coupling parameters are conveniently written as $J_{0}\pm\Delta_{0}/2$, where $J_{0}=\frac{E_{R}}{2}\left[ \frac{s_{1}\pi^2}{2}-\sqrt{s_{1}}-s_{1}\right] exp\left( -\frac{\sqrt{s_{1}}\pi^2}{4}\right)$
and $\Delta_{0}=s_{2}E_{R}exp\left( -\frac{\sqrt{s_{1}}\pi^2}{4}\right)$. The strength of the effective on-site interaction energy is $U_{eff}=U\int dz\,|w(z)|^4$. Here $U^{'}_{eff}=U_{eff}/V_{2d}$, $V_{2d}$ is the two dimensional area of radial confinement (i.e area of each pan cake shaped BEC). $\epsilon_{j}$ is the onsite energies and takes two distinct values ($\epsilon_{1}$ and $\epsilon_{2}$) corresponding to odd and even sites. In the mean field approximation, the operators $\hat{a}_{j}$ and $\hat{a}^{\dagger}_{j}$ are classical $c$ numbers, $\hat{a}_{j}=\phi_{j}$. Stationary states with a fixed total number of particles $N$ are obtained by requiring that the variation of $H-\mu N$ with respect to $\phi^{*}_{j}$ vanish. Here $\mu$ is the chemical potential. This yields the eigenvalues equation

\begin{equation}
\epsilon_{j}\phi_{j}+U_{eff}\left|\phi_{j}\right|^{2}\phi_{j}
-J_{j}\phi_{j+1}-J_{j-1}\phi_{j-1}-\mu\phi_{j}=0.
\end{equation}

We write $\phi_{j}$ as

\begin{equation}
\phi_{j}=g_{j}e^{ij2kd}.
\end{equation}

The eigenvalues are found as

\begin{equation}
\mu=\frac{2U_{eff}n_{0}
-\sqrt{\left[\Delta \epsilon \right]^{2}
+4\epsilon_{k}^{2}}}{2}.
\end{equation}

Where $\epsilon_{k}=\sqrt{4J_{0}^{2}\cos^{2}2kd+
\Delta_{0}^{2}\sin^{2}2kd}$ and $\Delta \epsilon=\epsilon_{1}-\epsilon_{2}$. The eigenvalue $\mu$ corresponds to the chemical potential for $k=0$.
The Bogoliubov spectrum of elementary excitation describes the energy of small perturbations with quasimomentum $q$ on top of a macroscopically populated state with quasi-momentum $k$. In the Bogoliubov approximation, we write the annihilation operator in terms of c-number part and a fluctuation operator as

\begin{equation}
\hat{a}_{j}=\left( \phi+\hat{\delta}_{j}\right) exp\left( -\frac{i\mu t}{\hbar}\right) 
\end{equation}

The resulting Bogoliubov equations for the fluctuation operator $\hat{\delta}_{j}$in the optical superlattice take the following form

\begin{equation}
i\hbar\dot{\hat{\delta}}_{j}=\left( 2U_{eff}n_{0}-\mu\right) \hat{\delta}_{j}-J_{j}\hat{\delta}_{j+1}-J_{j-1}\hat{\delta}_{j-1}+U_{eff}n_{0}\hat{\delta}_{j}^{\dagger}
\end{equation}

$n_{0}$ is the 2d average density of atoms per site of the lattice. The above equation is solved by constructing quasi-particles for the lattice, which diagonalize the Hamiltonian i.e

\begin{equation}
\hat{\delta}_{j}=\frac{1}{\sqrt{I}}\sum_{q}\left[u_{j}^{q}\hat{b}_{q}^{\dagger}e^{i(jq2d-\omega_{q}t)} -v_{j}^{q}\hat{b}_{q}e^{-i(jq2d-\omega_{q}t)}\right] 
\end{equation}

The quasi-particles obey the usual Bose-commutation relations

\begin{equation}
\left[\hat{b}_{q},\hat{b}_{q'}^{\dagger}\right]=\delta_{qq'}.
\end{equation}

The excitation amplitudes obey the periodic boundary conditions 

\begin{equation}
u_{j+1}^{q}=u_{j-1}^{q} , v_{j+1}^{q}=v_{j-1}^{q}
\end{equation}

Finally the phonon excitation frequencies are found to be
\begin{equation}
\hbar^{2}\omega_{q}^{2},_{\pm}
=\frac{1}{2}\left[(\beta_{1}^{2}+\beta_{2}^{2})
+2\epsilon_{q}^{2}-2U_{eff}^{2} n_{0}^{2}\right]
\pm \epsilon_{q} (\beta_{1}+\beta_{2})
\end{equation}

Where
\begin{equation}
\beta_{1}=U_{eff}n_{0}-\frac{\Delta \epsilon}{2}
+\frac{1}{2}\sqrt{(\Delta \epsilon)^{2}+16J_{0}^{2}}.
\end{equation}

\begin{equation}
\beta_{2}=U_{eff}n_{0}+\frac{\Delta \epsilon}{2}
+\frac{1}{2}\sqrt{(\Delta \epsilon)^{2}+16J_{0}^{2}}.
\end{equation}

\begin{equation}
\epsilon_{q}=\sqrt{4J_{0}^{2}\cos^{2}2qd+
\Delta_{0}^{2}\sin^{2}2qd}
\end{equation}

\begin{figure}[t]
\hspace{-1.5cm}
\includegraphics{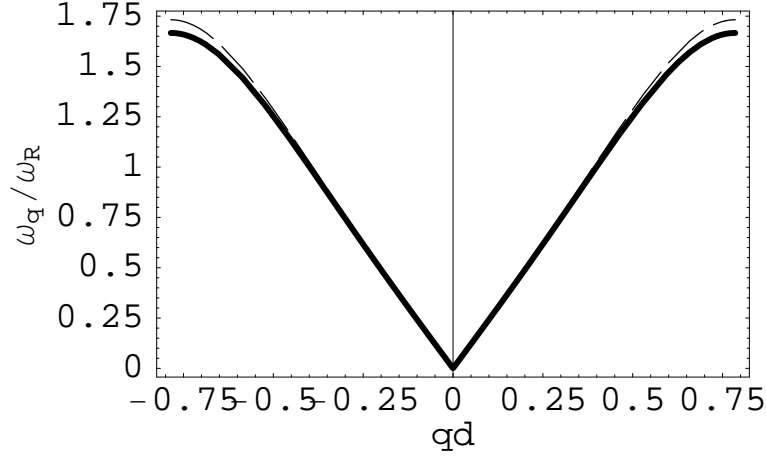} 
\caption{The acoustical branch of the Bogoliubov spectrum as a function of $qd$. $n_{0}U_{eff}/E_{R}=1$,$J_{0}/E_{R}=1$,$\Delta_{0}/E_{R}=0.1$. The bold curve is for $(\epsilon_{1}-\epsilon_{2})/E_{R}=1$, while the dashed curve is for $\epsilon_{1}=\epsilon_{2}$.The change in the Bogoliubov spectrum due to asymmetry is appreciable only near the band edge.}
\label{fig:figure_1_1}
\end{figure}

where $\hbar\omega_{q},_{-}$ is the acoustical branch. There is another branch called the gapped branch (analogue of the optical branch) whose energy is given by \cite{Bhattacherjee07} $\hbar\omega_{q},_{+}$.
In \figurename~\ref{fig:figure_1_1}, we find that for $\Delta \epsilon$ not large, the asymmetry due to $\phi\not=0$ does not appreciably change the Bogoluibov spectrum.It is only near the edge of the Brilliouin zone that a small change is visible.

Due to the above fact in the following we will only discuss the simple case when $\phi=0$ i.e experimentally $\phi$ does not deviate much from zero so that $\epsilon_{1}=\epsilon_{2}$. This case also allows us to tackle the problem analytically. The resulting equations for amplitudes and frequencies are solved to yield the Bogoliubov amplitudes as

\begin{equation}
|u_{j}^{q}|^{2}=|u_{j+1}^{q}|^{2}=\frac{1}{2}\left(\frac{\tilde{\epsilon}_{q,-}+n_{0}U_{eff}+\hbar \omega_{q,-}}{\hbar \omega_{q,-}} \right) 
\end{equation}

\begin{equation}
|v_{j}^{q}|^{2}=|v_{j+1}^{q}|^{2}=\frac{1}{2}\left(\frac{\tilde{\epsilon}_{q,-}+n_{0}U_{eff}-\hbar \omega_{q,-}}{\hbar \omega_{q,-}} \right) 
\end{equation}

\begin{equation}
u_{j}^{q}\,u_{j+1}^{*q}=\left(\frac{2J_{0}\cos2qd+i\Delta_{0}\sin2qd}{\sqrt{4J_{0}^2\cos2qd+\Delta_{0}^{2}\sin2qd}} \right)|u_{j}^{q}|^{2} 
\end{equation}

\begin{equation}
v_{j}^{q}\,v_{j+1}^{*q}=\left(\frac{2J_{0}\cos2qd+i\Delta_{0}\sin2qd}{\sqrt{4J_{0}^2\cos2qd+\Delta_{0}^{2}\sin2qd}} \right)|v_{j}^{q}|^{2} 
\end{equation}

\begin{equation}
v_{j}^{q}u_{j+1}^{q}=u_{j}^{q}v_{j+1}^{q}
\end{equation}
where $\hbar\omega_{q,-}=\sqrt{\tilde{\epsilon}_{q,-}(2n_{0}U_{eff}+\tilde{\epsilon}_{q,-})}$ and $\tilde{\epsilon}_{q,-}=2J_{0}-\sqrt{4J_{0}^{2}\cos^{2}2qd+\Delta_{0}^{2}\sin^{2}2qd}$.

\section{Superfluid Fraction and number fluctuations}

An interacting many body system is said to be superfluid, if a condensate exists. This happens when the one-body density matrix has exactly one macroscopic eigenvalue, which defines the number of particles in the condensate. The corresponding eigenvector describes the condensate wavefunction, $\psi_{0}(\vec{r})=e^{i\phi(\vec{r})}|\psi_{0}(\vec{r})|^{2}$. The superfluid velocity is given as

\begin{equation}
\vec{v}_{s}(\vec{r})=\frac{\hbar}{m^{*}}\vec{\nabla}\phi(\vec{r})
\end{equation}

Here $m^{*}$ is the effective mass of a single atom in the optical superlattice. We now write down the expression for the superfluid fraction based on the rigidity of the system under a twist of the condensate phase \cite{Rey03}. Suppose we impose a linear phase twist $\phi(\vec{r})=\frac{\theta z}{L}$, with a total twist angle $\theta$ over a length $L$ of the system (with ground state energy $E_{0}$) in the $z$ direction. The resulting ground state energy, $E_{\theta}$ will depend on the phase twist. Thus,

\begin{equation}
E_{\theta}-E_{0}=\frac{1}{2}m^{*}Nf_{s}v^{2}_{s}
\end{equation}

where $N$ is the total number of particles, $f_{s}$ is the superfluid fraction and $m^{*}=\frac{J_{0}\hbar^{2}}{2d^2 (4J_{o}^{2}-\Delta_{0}^{2})}$. Substituting equation (21) into (22)gives

\begin{equation}
f_{s}=\frac{4J_{0}(E_{\theta}-E_{0})}{N(4J_{0}^{2}-\Delta_{0}^{2})(\Delta \theta)^{2}}
\end{equation}

Here $\Delta \theta$ is the phase variation over $2d$. We now need to calculate the energy change $(E_{\theta}-E_{0})$ using second order perturbation theory, under the assumption that the phase change, $\Delta \theta$ is small. This yields

\begin{equation}
(E_{\theta}-E_{0})=\Delta E^{(1)}+\Delta E^{(2)}
\end{equation}
Where $\Delta E^{(1)}$ is the first order contribution to the energy change

\begin{equation}
\Delta E^{(1)}=-\frac{(\Delta \theta)^{2}}{2} \left\langle \psi_{0}|\hat T|\psi_{0}\right\rangle
\end{equation}

Here $|\psi_{0} \rangle$ is the ground state of the Bose-Hubbard Hamiltonian. The hopping operator $\hat T$ is given by

\begin{equation}
\hat T=-\sum_{j=1}^{I} J_{j} \left( \hat a^{\dagger}_{j+1} \hat a_{j}+\hat a_{j}^{\dagger}\hat a_{j+1}\right) 
\end{equation}

The second order contribution is written as 

\begin{equation}
\Delta E^{(2)}=-\left(\Delta \theta \right)^{2} \sum_{\nu \not= 0} \frac{|\left\langle \psi_{\nu}|\hat J|\psi_{0}\right\rangle|^{2}}{E_{\nu}-E_{0}}
\end{equation}

where the current operator $\hat J$ is 

\begin{equation}
\hat J=-\sum_{j=1}^{I} J_{j} \left( \hat a^{\dagger}_{j+1} \hat a_{j}-\hat a_{j}^{\dagger}\hat a_{j+1}\right) 
\end{equation}

The total superfluid fraction has two contributions.

\begin{equation}
f_{s}=f^{(1)}_{s}+f^{(2)}_{s}
\end{equation}

where

\begin{equation}
f^{(1)}_{s}=-\frac{2 J_{0}}{N(4J^{2}_{0}-\Delta^{2}_{0})} \left\langle \psi_{0}|\hat T|\psi_{0}\right\rangle
\end{equation}

\begin{equation}
f^{(2)}_{s}=\frac{2 J_{0}}{N(4J^{2}_{0}-\Delta^{2}_{0})}\sum_{\nu \not= 0} \frac{|\left\langle \psi_{\nu}|\hat J|\psi_{0}\right\rangle|^{2}}{E_{\nu}-E_{0}}
\end{equation}

Using the expressions for the various Bogoliubov amplitudes and frequencies, we can now evaluate $f_{s}^{(1)}$ and $f_{s}^{(2)}$.

\begin{equation}
f^{(1)}_{s}=\frac{2 J_{0}}{N(4J^{2}_{0}-\Delta^{2}_{0})} \sum_{j=1}^{I} J_{j} \left\langle \psi_{0}|\hat a^{\dagger}_{j+1} \hat a_{j}+\hat a_{j}^{\dagger}\hat a_{j+1}|\psi_{0}\right\rangle
\end{equation}

In the Bogoliubov approximation this takes the form

\begin{equation}
f^{(1)}_{s}=\frac{2 J_{0}}{N(4J^{2}_{0}-\Delta^{2}_{0})} \sum_{j=1}^{I} J_{j} \left\langle \psi_{0}|2\phi^{2}_{j}+\hat \delta^{\dagger}_{j+1} \hat \delta_{j}+\hat \delta_{j}^{\dagger}\hat \delta_{j+1}|\psi_{0}\right\rangle
\end{equation}

The fluctuation operators appearing in equation (32) are now written in terms of the quasi-particle operators.

\begin{eqnarray}
f^{(1)}_{s}=\frac{2J_{0}}{N(4J_{o}^{2}-\Delta_{0}^{2})}&&[\sum_{j=1}^{I} J_{j}(2\phi_{j}^{2})+\frac{1}{2}\sum_{j=1}^{I} \sum_{q,q'}J_{j} \left\langle\left[u_{j+1}^{q*}\hat{b}_{q}e^{iq(j+1)2d}-v_{j+1}^{q}\hat{b}_{q}^{+}e^{-iq(j+1)2d}\right] \left[u_{j}^{q'}\hat{b}_{q'}^{\dagger}e^{-iq'j2d}-v_{j}^{*q'}\hat{b}_{q'}^{\dagger}e^{iq'j2d}\right] \right\rangle \nonumber\\&& + \left\langle \left[u_{j}^{q*}\hat{b}_{q}^{\dagger}e^{-iqj2d}-v_{j}^{q}\hat{b}_{q}^{\dagger}e^{iqj2d}\right] \left[u_{j+1}^{q'}\hat{b}_{q'}^{\dagger}e^{iq'(j+1)2d}-v_{j+1}^{*q'}\hat{b}_{q'}^{\dagger}e^{-iq'(j+1)2d}\right]\right\rangle]  
\end{eqnarray}

Finally, we find in the zero temperature limit

\begin{equation}
f^{(1)}_{s}=\frac{4J_{0}}{N(4J_{o}^{2}-\Delta_{0}^{2})}\left\lbrace \sum_{j=1}^{I} J_{j}(\phi_{j}^{2})+\sum_{q}J_{0}\left( u_{2}^{*}u_{1}e^{i2qd}+u_{2}u_{1}^{*}e^{-i2qd}\right) \right\rbrace 
\end{equation}
 
Here, the summation runs over all quasi-momenta $q=\frac{\pi j}{Id}$ with $j=1,2,...(I-1)$. The normalization condition is obtained by putting $f_{s}^{(1)}=1$ when $d\rightarrow 0$.

\begin{equation}
\sum_{j=1}^{I} J_{j}(\phi_{j}^{2})+J_{0}\sum_{q}J_{0}2Re(u_{1}u_{2}^{*})=\frac{N(4J_{0}^{2}-\Delta_{0}^{2})}{4J_{0}}
\end{equation}

\begin{figure}[t]
\hspace{-1.5cm}
\includegraphics{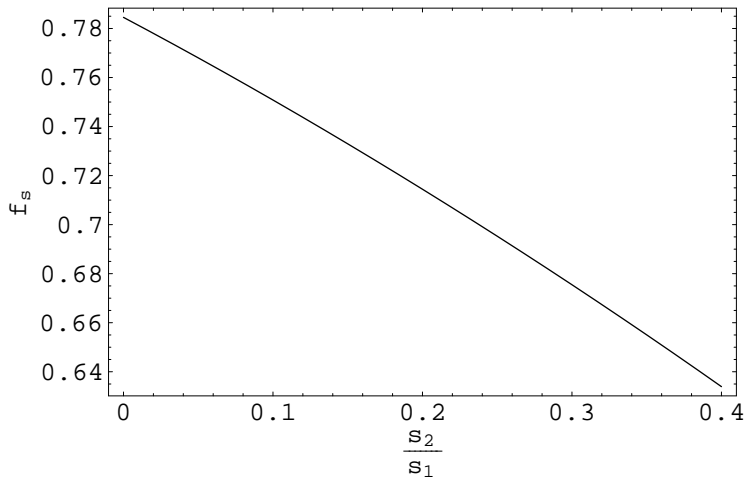} 
\caption{The superfluid fraction as a function of $s_{2}/s_{1}$ with $n_{0}U_{eff}/J_{0}=1$ , $I=10$ and $n=10$. As the strength of the secondary lattice increases with a fixed strength of the primary lattice, there is a quantum depletion of the condensate which is seen as a decrease in the superfluid fraction.}
\label{fig:figure_1}
\end{figure}

Using the Bogoliubov amplitudes derived in the previous section, one can show that $f_{s}^{(2)}=0$. Consequently, we find that the total superfluid fraction has contribution from just $f_{s}^{(1)}$. A plot (\figurename~\ref{fig:figure_1}) of the superfluid fraction as a function of $s_{2}/s_{1}$ reveals a decrease in the superfluid fraction as the strength of the secondary lattice increases. This is to be expected since in the presence of the secondary lattice,it has been shown that there exists a fractional filling Mott insulating state in the phase diagram \cite{Bounsante04}.This itself is an indication of a reduced superfluid fraction.This result is in accordance with earlier work of Rousseau et al. \cite{Marcos06} where they have considered the effect of a secondary lattice on an one dimensional hard core bosons(strongly correlated regime). As the strength of the secondary lattice increases, we approach the Mott-insulator transition. Since the phase twist is equivalent to the imposition of an acceleration on the lattice for a finite time, the condensate now in the superlattice seems to resist this acceleration or simply tries to resist the phase twist and thus there is a reduction in the superfluid flow. A direct consequence of the decrease of the superfluid fraction is a decrease in the number fluctuation, which we show below. 
Increasing the lattice depth reduces the tunneling rate between adjacent wells. This can be viewed as a reduction of the number fluctuations at each lattice site. As the probability of the atoms to hop between wells decreases, the number variance $\sigma_{n} $goes down. Quantum mechanically, this implies that the phase variance $\sigma_{\phi}$ describing the spread in relative phases between the lattice wells, has to increase. This effect can be seen directly by looking at the interference pattern of a BEC released from an optical trap. We can find an expression for the fluctuations in the relative number in each well as \cite{Rey03}

\begin{equation}
\left\langle \hat n_{i}^{2}-\left\langle \hat n_{i} \right\rangle^{2} \right\rangle=\frac{n}{I}\sum_{q}(u_{q}-v_{q})^{2} 
\end{equation}

and 

\begin{equation}
(u_{q}-v_{q})^{2}=\frac{\epsilon_{q}}{\hbar \omega_{q}}
\end{equation}

\begin{figure}[t]
\hspace{-1.5cm}
\includegraphics{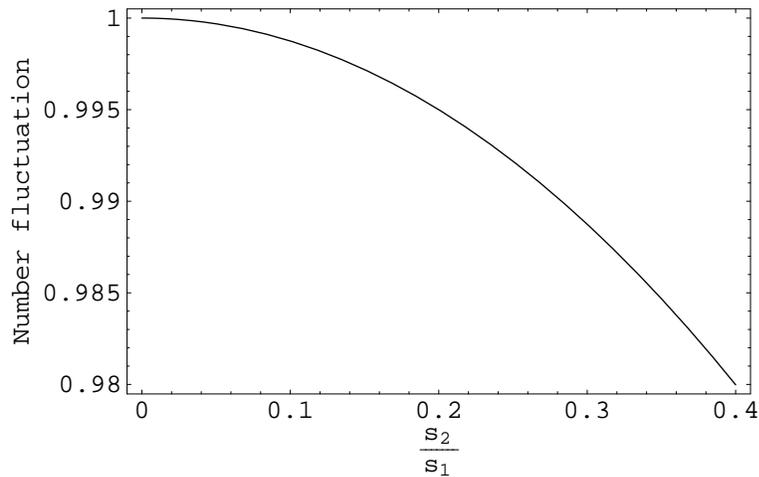} 
\caption{The number fluctuation as a function of $s_{2}/s_{1}$ with $n_{0}U_{eff}/J_{0}=1$ , $I=10$ and $n=10$. As the strength of the secondary lattice increases, there is a loss of superfluidity. The interplay of the interaction and tunneling terms renders number fluctuations energetically unfavorable. The number fluctuations decrease with increasing potential of the secondary lattice. There is a corresponding increase in the phase fluctuations.}
\label{fig:figure_2}
\end{figure}

$I$ is the total number of sites and $n$ is the mean number of atoms on each site of the lattice.A plot (\figurename~\ref{fig:figure_2})of the number fluctuations versus $s_{2}/s_{1}$ reveals as expected a decrease with increasing strength of the secondary lattice indicating a loss of phase coherence.  The number variance may be measured experimentally by studying the collapse $t_{c}$ and revival $t_{rev}$ times of the relative phase between sites \cite{Greiner03}. The relation is given by $\sigma_{n}=\frac{t_{rev}}{2\pi t_{c}}$. This reduction in the number fluctuation is also called as the atom number squeezing. This increased squeezing as a result of the secondary lattice has an important application in in improved atom interferometry since with increased squeezing the coherence time also increases \cite{Wei07}. These atom number squeezed states have reduced sensitivity to mean-field decay mechanisms. The secondary lattice then serves to coherently maintain a balance between coherence as well as the decoherence effects due to mean-field interaction.

\section{Dynamic structure factor}

The capability of the system to respond to an excitation probe transferring momentum $p$ and energy $\hbar \omega$ is described by the dynamic structure factor. In the presence of a periodic potential the dynamic structure factor takes the form

\begin{equation}
S(p,\omega)=\sum_{\alpha}Z_{\alpha}(p)\delta[\omega-\omega_{\alpha}(p)]
\end{equation}

where $Z_{\alpha}(p)$ are the excitation strengths relative to the $\alpha^{th}$ mode.$\alpha$ is the band label. For each value of the quasi-momentum $q$, there are infinite set of excitation energies $\hbar \omega_{\alpha}(q)$. It is often convenient to consider values of $q$ outside the first Brillouin zone and to treat the energy spectrum and Bogoliubov excitation amplitudes $u_{j,\alpha}^{q}$ and $v_{j,\alpha}^{q}$ as periodic with period $2q_{B}$. Here $q_{B}=\frac{\hbar \pi}{2d}$ is the Bragg momentum denoting the boundary of the first Brillouin zone. $p$ is assumed to be along the optical lattice (z axis), is not restricted to the first Brillouin zone since it is the momentum transferred by the external probe. 
The quantities $q$, $p$ and $q_{B}$ are related as $q=p+2lq_{B}$, $l$ is an integer. In the first Brillouin zone $l=0$. The excitation energies $\hbar \omega_{\alpha}(p)$ are periodic as a function of $p$ but this is not true for the excitation strengths $Z_{\alpha}$. The excitation strengths $Z_{\alpha}$ can be evaluated using the standard prescription \cite{Menotti02}

\begin{equation}
Z_{\alpha}(p)=|\int_{-d}^{d}\left[u_{\alpha}^{*q}(z)-u_{\alpha}^{*q}(z) \right]e^{ipz/\hbar}\phi(z) dz |^{2}
\end{equation}

Since $|u_{j,\alpha}^{q}|^{2}$=$|u_{j+1,\alpha}^{q}|^{2}$ and $|v_{j,\alpha}^{q}|^{2}$=$|v_{j+1,\alpha}^{q}|^{2}$, we will drop all $j$ dependence from the Bogoliubov amplitudes. The excitation frequencies for different $\alpha$ has already been derived in our earlier work .\cite{Bhattacherjee07} We are interested in the low energy region where $Z_{1}(p)$ is the dominating term arising from the first band. The dispersion law for the lowest band is 

\begin{equation}
\hbar \omega_{1}(p)=\sqrt{\tilde\epsilon_{p}(2n_{0}U_{eff}+\tilde\epsilon_{p})}
\end{equation}
 
\begin{equation}
\tilde\epsilon_{p}=2J_{0}-\sqrt{4J_{0}^{2}\cos^{2}{\left( \frac{2p\pi}{q_{B}}\right) }+\Delta_{0}^{2}\sin^{2}{\left( \frac{2p\pi}{q_{B}}\right)}}
\end{equation}

The behaviour of $Z_{1}(p)$ can be studies analytically in the tight binding limit. In this limit one can approximate the Bogoliubov amplitudes in the lowest mode as.

\begin{equation}
u_{\alpha}(z)=\sum_{j}e^{ij2qd/\hbar}f(z-2jd)
\end{equation}

and analogously for $v_{\alpha}(z)$, where $f(z)$ is a function localized near the bottom of the optical potential $V$ at $z=0$, and $j$ labels the potential wells. Within this approximation the function $f$ also characterizes the ground state order parameter which reads $\phi(z)=\sum_{j}f(z-2jd)$.
We can approximate the function $f(z)$ with the gaussian $f(z)=exp\left[-z^{2}/2\sigma^{2} \right]/\left(\pi^{1/4}\sqrt{\sigma}\right)$. The width $\sigma$ is found by minimizing the ground state energy

\begin{equation}
E_{0}=\frac{2}{2d}\int_{-d}^{d}\left[\frac{\hbar^{2}}{2m}|\frac{\partial \phi}{\partial z}|^{2}+\left\lbrace s_{1}E_{R}cos^{2}{\left( \frac{\pi z}{d}\right)}+s_{2}E_{R}cos^{2}{\left( \frac{\pi z}{2d}\right)} \right\rbrace|\phi|^{2}+\frac{U}{2}|\phi|^{4} \right]dz 
\end{equation}

and behaves like $\sigma\sim \frac{d}{(s_{1}+s_{2}/4)^{1/4}}$. After some trivial algebra we find

\begin{equation}
Z_{1}(p)=\frac{\tilde \epsilon_{p}}{\hbar \omega_{1}(p)}exp\left( {-\frac{\pi^2 \sigma^{2} p^{2}}{8d^{2}q_{B}^{2}}}\right) 
\end{equation}

\begin{figure}[t]
\hspace{-1.5cm}
\includegraphics{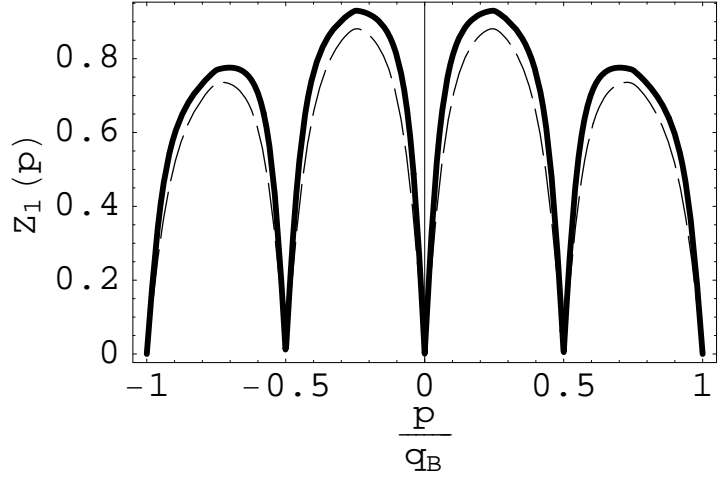} 
\caption{The excitation strength $Z_{1}(p)$ for two values of $\frac{s_{2}}{s_{1}}=0.1$ (solid line) and  $\frac{s_{2}}{s_{1}}=0.4$ (dashed line).$U_{eff}/J_{0}=0.2$. The figure shows both the oscillatory behaviour through $\frac{\tilde \epsilon (p)}{\hbar \omega_{1}(p)}$ and decaying behaviour at large $p$ through $exp\left( {-\frac{\pi^2 \sigma^{2} p^{2}}{8d^{2}q_{B}^{2}}}\right)$.On increasing the strength of the secondary lattice, $Z_{1}(p)$ is found to be quenched. The first maxima is found near the edge of the first Brillouin zone.}
\label{fig:figure_3}
\end{figure}

The expression for $Z_{1}(p)$ shows both the oscillatory behaviour through $\frac{\tilde \epsilon_{p}}{\hbar \omega_{1}(p)}$ and decaying behaviour at large $p$ through $exp\left( {-\frac{\pi^2 \sigma^{2} p^{2}}{8d^{2}q_{B}^{2}}}\right)$. \figurename~\ref{fig:figure_3} shows the excitation strength $Z_{1}(p)$ for two values of $\frac{s_{2}}{s_{1}}=0.1$ (solid line) and  $\frac{s_{2}}{s_{1}}=0.4$ (dashed line).On increasing the strength of the secondary lattice, $Z_{1}(p)$ is quenched. This behaviour can be understood by looking at the low $p$ limit of $S(p)=\int S(p,\omega) d\omega=\frac{|p|}{2\sqrt{m^*n_{0}U_{eff}}}$. on increasing $s_{2}$, $m^*$ increases and hence $S(p)$ decreases. The presence of the secondary lattice results in the suppression of $Z_{1}(p)$. The system now becomes more heavy and is not able to respond to an external excitation probe. The momentum transferred is now comparatively less. Note that in the absence of interactions, the oscillatory behaviour disappears and the strength reduces to $Z_{1}(p)= exp\left( {-\frac{\pi^2 \sigma^{2} p^{2}}{8d^{2}q_{B}^{2}}}\right)$. This shows that the effect of the secondary lattice on the quenching is present only in the presence of interactions.The zeroes of $Z_{1}(p)$ at $p=2lq_{B}$ reflects the phonon behaviour of the excitation spectrum which also vanishes at the same values. The quantity $Z_{1}(p)$ can be measured in Bragg spectroscopy experiments by applying an additional moving optical potential in the form of $V_{B}(t)=V_{0}\cos{(\frac{pz}{\hbar})-\omega t}$. The momentum and the energy transferred by the Bragg pulse must be tuned to the values of $p$ and $\hbar \omega$ corresponding to the first Bogoliubov band.

\section{Quasimomentum distribution of the Mott insulator in an optical superlattice: visibility of fringes}

For a Bose-Einstein condensate released from an optical lattice, the density distribution after expansion shows a sharp interference pattern. In a perfect Mott-insulator, where atomic interactions pin the density to precisely an integer number of atoms per site, phase coherence is completely lost and no interference pattern is expected. The transition between these two limiting cases happens continuously as the lattice depth is increased. In this section, we will look into the influence of increasing the strength of the secondary lattice on the phase coherence of the insulating phase. We consider an integer number $n$ of atoms per site and $J_{0}\pm \frac{\Delta_{0}}{2}<< U_{eff}$. In this situation the gas is in the Mott-insulator phase. The Mott insulating phase has the property that the fluctuations in the average number of particles per site goes to zero at zero temperature. These fluctuations can be described as quasihole and quasiparticle excitations. To calculate the quasimomentum distribution $S(k)$ for a finite tunneling, path integral techniques can be applied to obtain the single-particle Green function, $G(\vec{k},\omega)$. The quasi-momentum distribution is an useful quantity to describe the interference pattern observed after release of the cold cloud from the optical lattice. From the absorption image of such an interference pattern, the phase coherence of the atomic sample can be directly probed.To extract quantitative information from time-of-flight absorption images, one can use the usual definition of the visibility of interference fringes \cite{Gerbier06},

\begin{equation}
V=\frac{S_{max}-S_{min}}{S_{max}+S_{min}}
\end{equation}
 
The quasimomentum distribution $S(k)$ contains information about the many-body system which is periodic with the periodicity of the reciprocal lattice corresponding to the secondary lattice. Thus to predict the interference pattern in the superlattice, our goal is to calculate $S(k)$ as function of $J_{0}$ and $\Delta_{0}$. 
We calculate the quasiparticle and quasihole dispersions using the functional integral formalism of Van Oosten et. al. \cite{Oosten01}. The grand-canonical partition function in terms of the complex functions $a_{j}^{*}(\tau)$ and $a_{j}(\tau)$ is written as

\begin{equation}
Z=Tre^{-\beta H}=\int Da^{*}\, Da\, exp\left\lbrace -S\left[ {a^{*},a}\right]/ \hbar \right\rbrace 
\end{equation}

where the action $S[a^{*},a]$ is given by

\begin{equation}
S[a^{*},a]=\int_{0}^{\hbar \beta} d\tau \left[\sum_{j}a_{j}^{*}\left( \hbar \frac{\partial}{\partial \tau}-\mu\right) a_{j}-\sum_{j,j'}J_{jj'}a_{j}^{*}a_{j'}+\frac{U_{eff}}{2}\sum_{j}a_{j}^{*}a_{j}^{*}a_{j}a_{j} \right] 
\end{equation}

$J_{j,j'}$ is the hopping element, $\beta=1/k_{B}T$, $k_{B}$ is the Boltzmann constant and $T$ is the temperature. A Hubbard-Stratonovich transformation decouples the hopping term.

\begin{equation}
S\left[a^{*},a,\psi ^{*},\psi \right]=S\left[ a^{*},a\right]+\int_{0}^{\hbar \beta}d\tau \sum_{j,j'}\left(\psi_{j}^{*}-a_{j}^{*} \right)J_{jj'}\left(\psi_{j}-a_{j}\right)    
\end{equation}

Here $\psi^*$ and $\psi$ are the order parameter fields. Integrating over the original fields $a_{j}^{*}$ and $a_{j}$, we find

\begin{eqnarray}
exp\left(-S^{eff}\left[\psi^{*},\psi \right]/\hbar  \right)= && exp\left(-\frac{1}{\hbar}\int_{0}^{\hbar \beta}d\tau \sum_{j,j'}J_{jj'}\psi_{j}^{*}\psi_{j'} \right)\int Da^{*}\,Da\, exp\left(-S^{(0)}[a^{*},a]/\hbar \right)\nonumber\\ && exp\left[ -\frac{1}{\hbar}\int_{0}^{\hbar \beta}d\tau\left( -\sum_{j,j'}J_{jj'}\left( a_{j}^{*}\psi_{j'}+\psi_{j}^{*}a_{j'}\right) \right) \right]  
\end{eqnarray}

Here $S^{(0)}[a^{*},a]$ is the action for $J_{j,j'}=0$. We can now calculate $S^{eff}$ perturbatively by Taylor expanding the exponent in the integrand of equation (49) and find the quadratic part of the effective action using $\left\langle a_{j}^{*}a_{j'}^{*} \right\rangle _{S^{(0)}}=\left\langle a_{j}a_{j'} \right\rangle _{S^{(0)}}=0$, $\left\langle a_{j}^{*}a_{j'} \right\rangle _{S^{(0)}}=\left\langle a_{j}a_{j'}^{*} \right\rangle _{S^{(0)}}=\left\langle a_{j}a_{j}^{*} \right\rangle _{S^{(0)}}\delta_{jj'}$, 

\begin{equation}
S^{(2)}[\psi^{*},\psi]=\int_{0}^{\hbar \beta}d\tau\, \left( \sum_{j,j'}\psi_{j}^{*}(\tau)\psi_{j'}(\tau)-\frac{1}{\hbar}\int_{0}^{\hbar \omega}d\tau '\sum_{jj'ii'}J_{jj'}J_{ii'}\psi_{j'}^{*}(\tau)\left\langle \ a_{j}(\tau)a_{i}^{*}(\tau ')\right\rangle_{S^{(0)}}\psi_{i'}(\tau ') \right) 
\end{equation}

We first evaluate the part linear in $J_{jj'}$ for nearest neighbours. We have

\begin{equation}
\sum_{j,j'}\psi{j}^{*}(\tau)\psi_{j'}(\tau)=\left(J_{0}+\frac{\Delta_{0}}{2} \right)\psi_{j}^{*}\psi_{j+1}+ \left(J_{0}-\frac{\Delta_{0}}{2} \right)\psi_{j}^{*}\psi_{j-1}
\end{equation}

We now introduce $\psi_{j}=[u_{k}+i(-1)^{j}v_{k}]exp(ij2kd)$.  As the condensate moves from one well to the next, it acquires an additional phase, which depends on the height of the barrier. As the height alternates and hence the tunneling parameter, the phase also alternates. This picture is conveniently represented by the $j$ dependent amplitude. This implies

\begin{eqnarray}
\sum_{j,j'}\psi{j}^{*}(\tau)\psi_{j'}(\tau)=&&2J_{0}\left[ |u_{k}|^{2}-|v_{k}|^{2}\right] \cos(2kd)-i2J_{0}\left[ u_{k}v_{k}^{*}+u_{k}^{*}v_{k}\right] \cos(2kd)+i\Delta_{0}\left[ |u_{k}|^{2}-|v_{k}|^{2}\right] \sin(2kd)\nonumber\\&&+\Delta_{0}\left[ u_{k}v_{k}^{*}+u_{k}^{*}v_{k}\right] \sin(2kd)
\end{eqnarray}

For the imaginary part to vanish we have for the one-dimensional optical lattice

\begin{equation}
u_{k}^{*}v_{k}=u_{k}v_{k}^{*}=\psi_{k}^{*}\psi_{k}\frac{\Delta_{0}\sin(2kd)}{2\epsilon_{k}}
\end{equation}

\begin{equation}
|u_{k}|^{2}-|v_{k}|^{2}=\psi_{k}^{*}\psi_{k}\frac{2 \Delta_{0}\cos(2kd)}{\epsilon_{k}}
\end{equation}

\begin{equation}
\epsilon_{k}=\sqrt{4J_{0}^{2}\cos^{2}(2kd)+\Delta_{0}^{2}\sin^{2}(2kd)}
\end{equation}

Finally we have,

\begin{equation}
\sum_{j,j'}\psi_{j}^{*}(\tau)\psi_{j'}(\tau)=\sum_{k}\epsilon_{k}\psi_{k}(\tau)\psi_{k}^{*}(\tau)
\end{equation}

Next we calculate the part that is quadratic in $J_{j,j'}$. We can treat this part by looking at double jumps.

\begin{eqnarray}
&& \sum_{j'ii'}J_{jj'}J_{ii'}\psi_{j'}^{*}(\tau)\left\langle a_{j}(\tau)a_{i}^{*}(\tau ') \right\rangle_{S^{(0)}}\psi_{i'}(\tau ')= \left\langle a_{j}(\tau)a_{j}^{*}(\tau ') \right\rangle_{S^{(0)}}\sum_{j'i'}J_{jj'}J_{ji'}\psi_{j'}^{*}(\tau)\psi_{i'}(\tau ')\nonumber\\&& = \left\langle a_{j}(\tau)a_{j}^{*}(\tau ') \right\rangle_{S^{(0)}}\left\lbrace\sum_{j'j'}J_{jj'}J_{jj'}\psi_{j'}^{*}(\tau)\psi_{j'}(\tau ')+J_{jj'}J_{jj'\pm 2}\psi_{j'}^{*}(\tau)\psi_{j'\pm 2}(\tau ') \right\rbrace 
\end{eqnarray}

The first term in the summation is a jump forward, followed by a jump backward. The second is two jumps in the same direction. The above quadratic term then reduces to

\begin{equation}
\sum_{j'ii'}J_{jj'}J_{ii'}\psi_{j'}^{*}(\tau)\left\langle a_{j}(\tau)a_{i}^{*}(\tau ') \right\rangle_{S^{(0)}}\psi_{i'}(\tau ')=\left\langle a_{j}(\tau)a_{j}^{*}(\tau ') \right\rangle_{S^{(0)}} \sum_{k} \epsilon_{k}^{2}\psi_{k}^{*}(\tau)\psi_{k}(\tau ')
\end{equation}

The Green's function is then easily calculated by following the steps indicated in ref.\cite{Oosten01}

\begin{equation}
\frac{G(\vec k,\omega)}{\hbar}=\frac{Z_{k}}{\hbar \omega+\mu-E_{k}^{(+)}}+\frac{1-Z_{k}}{\hbar \omega+\mu-E_{k}^{(-)}}
\end{equation}

The quasiparticle energies $E_{k}^{\pm}$ are derived as

\begin{equation}
E_{k}^{\pm}=-\frac{\epsilon_{k}}{2}+U_{eff}\left(n-\frac{1}{2} \right)\pm \frac{1}{2}\sqrt{\epsilon_{k}^{2}-4\epsilon_{k}U_{eff}\left(n+\frac{1}{2}\right) +U_{eff}^{2} } 
\end{equation}

The particle weight $Z_{k}$ is 

\begin{equation}
Z_{k}=\frac{\left(E_{k}^{(+)}+U_{eff} \right) }{\sqrt{\epsilon_{k}^{2}-4\epsilon_{k}U_{eff}\left(n+\frac{1}{2}\right) +U_{eff}^{2} } }
\end{equation}

The quasimomentum distribution can be directly calculated from the Green function $G(\vec k,\omega)$ using the relation

\begin{equation}
S(\vec k)=-i\lim_{\delta t\to 0}\int \frac{d\omega}{2\pi}G(\vec k,\omega)exp\left(-i\omega \delta t \right) 
\end{equation}

This yields

\begin{equation}
S(\vec k)=n\left(\frac{-\frac{\epsilon_{k}}{2}+U_{eff}\left(n+\frac{1}{2} \right)}{\sqrt{\epsilon_{k}^{2}-4\epsilon_{k}U_{eff}\left(n+\frac{1}{2}\right) +U_{eff}^{2} }}-\frac{1}{2} \right) 
\end{equation}

\begin{figure}[t]
\hspace{-1.5cm}
\includegraphics{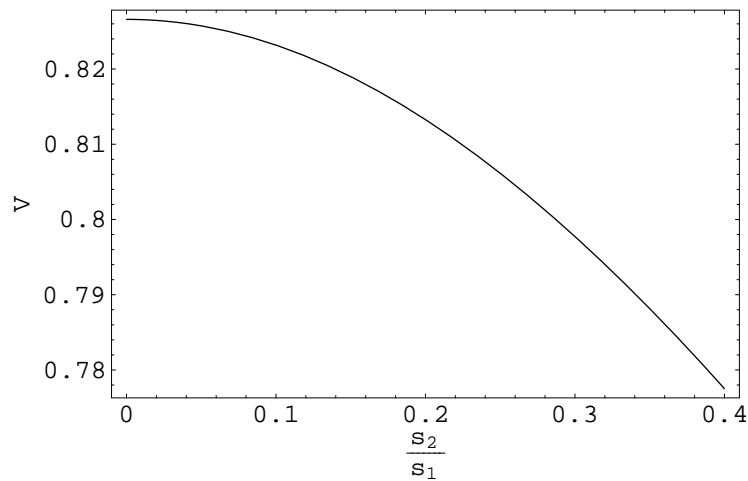} 
\caption{The visibility of the interference pattern produced by an ultracold cloud released from an optical superlattice as a function of $s_{2}/s_{1}$ with $U_{eff}/J_{0}=40$ and $n_{0}=3$. As the strength of the secondary lattices increases, the visibility worsens since the system gradually goes deeper into the Mott insulator regime and a corresponding gradual loss of long range coherence. A finite visibility even for a Mott-insulator is due to short range coherence since the system consists of a small admixture of particle-hole pairs on top of a perfect Mott-insulator. A loss of visibility in the superlattice naturally means that there is loss of particle-hole pairs.}
\label{fig:figure_4}
\end{figure}

$S(\vec k)$ is simply the quasi-momentum distribution which tells us about the many-body system. The visibility of the interference pattern of a cloud of BEC released from an optical superlattice as a function of the strength of the secondary lattice is shown in figure 4. As the strength of the secondary lattices increases, the visibility worsens since the system gradually goes deeper into the Mott insulator regime and a corresponding gradual loss of long range coherence. A finite visibility even for a Mott-insulator is due to short range coherence since the system consists of a small admixture of particle-hole pairs on top of a perfect Mott-insulator. A loss of visibility in the superlattice naturally means that there is loss of particle-hole pairs.

\section{Conclusions}
We have studied the effect of a one dimensional optical superlattice on the superfluid fraction, number squeezing, dynamic structure factor and the quasi-momentum distribution of the Mott-insulator. We have shown that the secondary lattice suppresses the superfluidity due to quantum depletion of the condensate and hence generates atom-number squeezed state which offers a possibility to create states with reduced sensitivity to mean field decay mechanism useful for improved atom-interferometry. A coherent control over the phase coherence in the superfluid as well as the Mott-insulating state can be achieved which has important applications in quantum computing.

\begin{acknowledgments}
The author is grateful to the Max Planck Institute for Physics of Complex Systems, Dresden, Germany for the hospitality and for providing the facilities for carrying out the present work.
\end{acknowledgments}

\end{document}